\documentstyle[12pt,aps,preprint,epsf,tighten]{revtex}

\begin{document}
\draft
\title{
 Plateau in Above-Threshold-Ionization Spectra and Chaotic Behavior in Rescattering Process
}
\author{
Jie Liu $\,^{1,2}$,     Shi-gang Chen $\,^2$, and 
Bambi Hu $\,^{1,3}$ 
 }

\address{$^1$Department of Physics and Center for Nonlinear Studies  ,
\\ Hong Kong Baptist University, Hong Kong\\
$^2$Institute of Applied Physics and
Computational Mathematics,\\P.O.Box.8009,  100088 Beijing, China\\
$^3$Department of Physics, University of Houston, Houston,TX 77204 USA\\
}
\maketitle
\date{}
\begin{abstract}
An improved quasistatic model is used to describe 
the ionization process of atoms in intense linearly polarized fields.
Numerical calculations of Above-Threshold-Ionization (ATI) energy spectra
and photoelectron angular distributions (PAD)  of hydrogen atoms
 reveal clearly plateau and side lobe structures,
 which are in  good agreement with recent  ATI experiments.
Our results  show that, the existence of this  plateau and side lobes is
consequences of the classical kinematics of the electrons in combined
atom and laser fields. Furthermore, we find that  the onset of these unusual
phenomena is related to the onset of the chaotic behavior in
the   system.
\end{abstract}
\section{Introduction}
The most recent important finding in Above-Threshold-Ionization (ATI) experiments is 
the existence of  a  plateau formed by the high-order ATI peaks which halts and even
temporarily reverses the decrease of the heights of the peaks with
increasing order$[1]$. Moreover, the Photoelectron Angular Distribution (PAD) in the
transition region at the onset of the plateau, seem unusual: while
the angular distributions  both below and well above this region
are strongly concentrated in the field direction, additional 'side lobes'
at angles between $\pi/6$ and $\pi/4$ with respect to the field direction
develop in the transition region $[1]$.

It has been concluded by numerical calculation of the one-electron Schroedinger
 equation that the new phenomena can be understood in the context of single-
 electron ionization dynamics $[1]$. The conjecture that the electron
 returning to the ion core will cause a number of observable consequences
 , such as the cut-off law of the high-order harmonic production, has been made
 by many
 authors $[2,3]$ and it now seems evident that recent findings of ATI result
 from
 the rescattering effect as well. A fully quantum mechanical calculations
  for
 the rescattering effect based on a two-step model with a simplified delta
 source has been well performed by Becker et al $[4]$. An extensive analysis
 of this effect based on the Keldysh-Faisal-Reiss (KFR) theory with realistic
 atomic potential has also been presented $[5]$. Lewenstein and coworkers
have made analysis on lobes in ATI and its relation to rescattering by 
using semiclassical method $[6]$. However, 
  the precise physical origin of the plateau and
 lobes is still unclear. Are they due to some genuinely quantum mechanical effect such as
 interference or can they be traced back to the essentially classical
 behavior of the electrons in the combined atom and laser fields?
 This question
 is partially answered by  Paulus {\it et al} $[7]$. However 
  the emission of photoelectrons
 in the direction of the laser electric fields  is strongly
 underestimated by their simple classical model.

In this paper, first we shall develope a 3D quasistatic model which
generalizes the well-known quasistatic model$[2,3]$ by including the 
effect of Coulomb potential  on  electron motion after tunnel-ionization.
 Taking the simple Hydrogen atom as an example, we shall investigate
 its ionization process in an intense laser field based on
 the improved model. Our  results  show clearly that,
 the  plateau structure in
 ATI spectra and side lobes in the PADs can be
 attributed to the pure classical behavior of electrons in the combined
 Coulomb and laser field.
  Second, we shall present numerical evidence that, 
 the unusual phenomena observed in ATI data result from  a kind of  irregular
 motion in the nonlinear dynamical system. 
   Our discussions connect the optical field ionization problem
  with another active field, namely, Chaotic Dynamics. This connection
  will  enrich and  deepen our understandings on atom-laser interaction.

\section{An Improved Quasistatic Model}
As Keldysh parameter $\gamma \equiv (I_0/2U_p)^{1/2}\ll 1, \hbar \omega /2U_p\ll 1$ and field
strength $F < F_{th}$,  tunneling
ionization occurs. Here, $\hbar$ is the Planck constant, 
$I_0$ is the ionization potential of the electron, $\omega$ is the field frequency,
 $Z$  
the core charge, 
 $U_p=\frac{Z^2e^2F^2}{4m_e\omega^2}$
 the ponderomotive potential, $F_{th}$  the threshold field (its meaning and value will be given in following
disscussions). 
The probability of the tunnel-ionization 
 is first obtained by Landau $[8]$ 
for  a hydrogen atom ( in the
ground state) 
and then be  extended
to  complex atoms $[9]$.
 More recently,
Delone and Krainov derived an analytical  expression for the
probabilities of tunnel-ionization of atoms and for the energy and
angular electron spectra in a strong low-frequency electric field $[10]$.
Their 
discussions give a perfect description  of the tunneling problem and
will be  helpful to provide the initial conditions of an electron after
tunneling in our 3D model.

In the  second step  of the quasistatic procedure,
by including the Coulomb potential as well as the laser fields, we obtain 
 a complete Newtonian equation,
which  describes the motion  of an electron after tunneling ionization. 
\begin{equation}
m_e \ddot{\vec{r}} = -\frac{Ze^2\vec{r}}{r^3} + (-ZeF cos\omega t \vec{e_z}
+\beta ZeF sin\omega t \vec{e_y}) \,\,\, ,
\end{equation}
where  $\beta =0$ for linearly polarized laser
and $\beta=\pm 1$ for circular polarization.

It is convenient to introduce the compensated energy $E_c$ advocated by
Leopold and Percival$[11]$ (the $E_c$ is just the Hamiltonian function in the velocity gauge),  
\begin{equation}
E_c = \frac{m_e}{2}[\dot{\vec{r}}+(ZeF/m_e\omega)\sin(\omega t) \vec{e_z}
-\beta (ZeF/m_e\omega)\cos(\omega t) \vec{e_y}
                                  ]^2
                                  -Ze^2/r \,\,\, .
\end{equation}
When an electron is ionized completely,
the Coulomb potential is weak enough and $E_c$ tends to be a
positive constant  which is just the ATI energy ($E_{ATI}$) in
ultrashort pulse laser.

In what follows,  taking the simple Hydrogen atom   in the
linearly polarized fields as an example we shall   demonstrate
this approach. 
For simplicity,
the atomic units are  used ( that is, $m_e = e = \hbar = 1 a.u.$). 
 The field parameters are chosen as $F=0.06 a.u. (1.2\times  10^{14} w/cm^2) $ and
$\omega=0.04242 (\lambda=1064nm)$, $I_0=U_p= 0.5 a.u. (13.5 eV)$, then
$\gamma = 0.71 < 1$ and $\hbar\omega/2U_p=0.04242 <<1$, and therefore
this situation
is well in the tunnel ionization regime.

We first recall the deduction of the effective potential in $[8]$ for the self-consistence of the paper.
Schroedinger's equation for the hydrogen atom in a uniform 
electric field  is of the form
\begin{equation}
(\frac 1 2 \Delta + I_0 + \frac 1 r - \epsilon z) \Psi = 0,
\end{equation}
where $\Delta$ is the Laplace operator; $\epsilon
=F\cos\phi_0$, 
 the quasistatic field;
$\phi_0 = \omega t_0$, the field phase at time $t_0$.

Let us introduce the parabolic coordinates $\xi, \eta, \psi$
\begin{equation}
x=\sqrt{\xi\eta}\cos\psi , y=\sqrt{\xi\eta}\sin\psi, z=(\xi-\eta)/2.
\end{equation}
We seek the eigenfunctions in the form
\begin{equation}
\Psi = \frac{f_1(\xi)f_2(\eta)}{\sqrt{\xi\eta}} e^{i m \psi},
\end{equation}
where $m$ is the magnetic quantum number. 

Then equation (3) is brought into the form 
\begin{equation}
\frac{d^2f_1}{d\xi^2} + ( \frac {I_0}{2} +\frac{\beta_1}{\xi} -
\frac{m^2-1}{4\xi^2} -\frac 1 4 \epsilon \xi) f_1 =0,
\end{equation}
\begin{equation}
\frac{d^2f_2}{d\eta^2} + ( \frac{I_0}{2} +\frac{\beta_2}{\eta} -
\frac{m^2-1}{4\eta^2} +\frac 1 4 \epsilon \eta) f_2 =0,
\end{equation}
\begin{equation}
\beta_1 + \beta_2 =1.
\end{equation}
 We shall regard the energy $I_0$ as a parameter which has a definitive 
 value. Solving the equations, as functions of $I_0$ and $\epsilon$, and 
 then the condition $\beta_1 + \beta_2 = 1 $ will give the energy as 
 a function of the external field.

 Above equations show that there exist a potential barrier along
 the $\eta$ coordinate, the ionization of the electron from the
 atom in the direction $z \to -\infty$ corresponds to its passage
  into the region of large $\eta$. To determine the ionization
   probability it is necessary to investigate the form of the wave 
   function for large $\eta$ and small $\xi$. Neglecting the stark
    shift we have $I_0 = -\frac 1 2$ and $\beta_1 =\beta_2 = \frac 1 2 $
    and $m = 0$. 
    In the absence of the field, the wave function is
    \begin{equation}
    \Psi_0  = e^{-(\xi+\eta)/2}/\sqrt{\pi}.
    \end{equation}
    When the field is present, the dependence of the $\Psi$
    on $\xi$ can be regarded as being the same as (9), while to determine
    its dependence on $\eta$ we have the equation (from (7)),
\begin{equation}
\frac{d^2\Phi}{d\eta^2} + ( -\frac 1 4 +\frac{1}{2\eta} +
\frac{1}{4\eta^2} +\frac 1 4 \epsilon \eta) \Phi =0,
\end{equation}
where $\Phi = \sqrt{\eta}\Psi$.

Above equation  has the form of the one-dimensional Schrodinger's equation
with the potential
 $U(\eta) = -\frac{1}{4\eta} - \frac{1}{8\eta^2} -
\frac 1 8 \epsilon \eta$
 and the energy $K = -\frac 1 8$. 
 The turning point
( at the outer edge of the suppressed  potential )
where an electron born at time $t_0$ is determined by $U(\eta)=K$ and
expressed by
$$\eta_0=\frac{1}{3\epsilon}+2s^{-3/2}\cos\theta, s=\sqrt{-(\frac p 3)^3},
\theta=\frac 1 3 \arccos(-\frac{q}{2s}),$$
\begin{equation}
 p=-\frac{1}{3\epsilon^2}+\frac{2}{\epsilon},
q=-\frac{2}{9\epsilon^3}+\frac{2}{3\epsilon^2}+\frac{1}{\epsilon}
\,\,\, .
\end{equation}
As  $\epsilon > F_{th}$, the turning point
will be
 complex, which determines the threshold value
 of the field $F_{th}= 0.113 a.u.$.

 WKB method is used to  obtain the tunneling rate of the above system
and gives$[8]$
$
 w(0)=\frac{4}{\epsilon}
 exp(-\frac{2}{3\epsilon
})$, which is  used in  the original quasistatic model$[2,3]$ (quasi one-dimensional model). This model can qualitatively predict the ATI spectrum and provide
a analytic expression of the initial parameters in the problem of 
the optical field
 ionization X-ray laser.
However, this simple model can not include the rescattering effects that result
in many important observable phenomena, such as Plateau in ATI spectra. 
Morever, it can not predict the PADs for it is an one-dimensional model.
Recently, Delone {\it et al} generalized the above tunneling formula and
obtained the probabilities of tunnel ionization for both energy and angular 
energy as follows$[10]$,
\begin{equation}
w(p_{\bot})=w(0)w(1), w(1)=exp(-p_{\bot}^2/\epsilon)
\end{equation}
where $p_{\bot}$ is the perpendicular momentum of the photoelectrons.

The above formula will  be used  to weight an classical trajectory 
in our three-dimensional quasistatic model.

Since the system is azimuthal symmetric about the polarization axis,
we can restrict the motion
of an electron on the
plane $(x,z)$. In  addition, for the exponential decay of the tunneling
rate  along the $\xi$ axis, it is a reasonable assumption that $\xi_0=0$ for 
an electron just after tunneling.  Thus,     the initial position
of an electron born at time $t_0$ is given by $x_0=0, z_0=-\eta_0/2$ from equation (4). The initial
velocity is set to be $v_{z}=0, v_{x}= v_{x0}$.
The weight of each trajectory
is evaluated  by the Delone's expression of tunnel rate (12),
\begin{equation}
w(\phi_0,v_{x0}) = w(0) \bar w(1),
 w(0)=\frac{4}{\epsilon}
 exp(-\frac{2}{3\epsilon
}),
\bar w(1)=\frac{v_{x0}}{\epsilon\pi} exp(-v_{x0}^2/\epsilon) \,\,\, .
\end{equation}

In applying Delone's formula to our problem, the second term should
be normalized to present form. From this formula, one can readily verify that, for a fixed initial
perpendular velocity $v_{x0}$, the trajectory corresponding to the initial field phase $\phi_0=0$
(i.e. at maximum quasistatic field $\epsilon = F$) has the maximum probability; for a fixed initial
field phase $\phi_0$, the trajectories corresponding to $v_{x0} = \sqrt{\epsilon/2}$ has 
the maximum probability.

In our computations, $10^5$ initial points are randomly distributed in the parameter plane
$-\pi/2<\phi_0<\pi/2, v_{x0}>0$ so that the weight of the
chosen trajectory   is larger than $10^{-11}$. Each trajectory is traced
for  such a long time that  the electron is actually ionized
(this can be evaluated by using the compensated energy $E_{ATI}$.)
The  final ATI spectra and the PADs can be obtained by
making statistics on an ensemble of  classical trajectories. The results
have been tested for numerical convergence by increasing the number of
trajectories.

Before we turn to our numerical results, we would like to mention that similar
model has been used successifully to double ionization of He atoms by Brabec et al $[12]$. Differently, in this model, the initial conditions of electrons 
after tunneling is derived strictly from the effective potential in Landau's 
tunneling theory rather than Coulomb potential. This deduction gives the
 correct threshold field $F_{th} = 0.113 a.u$ for 3d hydrogen atom. This 
model is restricted to the situation where the external field strength is 
smaller than the threshold value.
\section{Results}
\subsection{ATI Spectra and PADs of Photoelectrons}
In Fig.1, we demonstrate the ATI spectra and the total angular
 distribution (of the emitted electrons with respect to the angle $\theta$,
i.e. at the detector) calculated from our model.
  The original simple  quasistatic model predicts
 an sharply decreasing ATI spectra curve  and the photoelectron energy can
 not be larger than $2U_p$$[2,3]$. Our results show clearly that the rescattering
 increases the fraction of the hot electrons. This is due to
 that an electron has a higher  probability of staying in the vicinity of the
 nucleus and then absorb more photons in the  rescattering process. In particular,
 the  ATI spectrum exhibits  a sharply decreasing slope (region I, 0 - 2$U_p$) followed by
 a plateau (the  transition region II, 2$U_p$ - 8$U_p$)
  and again a sharply decreasing slope (region III, $>$ 8$U_p$). The height of the plateau is
 three orders of magnitude below the maximum of the spectrum and its width
 is about $6U_p$. This phenomenon is qualitatively in agreement with
 the existing experiments and various theories $[1,4,5,6]$.
 The distinctive feature is the plateau's extension about $6U_p$ width 
 to rather high energy before abruptly delcining at about $8U_p$.
This  differs  from the previous experiments $[1]$ but is much closer to
recent experiment $[13]$
which is performed in strong tunneling limit.
 The total angular distribution in Fig.1b  contains only the electron
 initiated in phase interval $[-\pi/2, \pi/2]$. The angular distribution
 of the electrons originated in $[\pi/2,3\pi/2]$ will be the mirror image
 with respect to $90^0$, so  the sum of the two contributions will show
 a main concentration in the field direction.

 Furthermore, in Fig.2 we calculate   statistics on the angular distribution
 of photoelectrons in three different energy regions.
 The most striking  feature of the plots is the  existence of a
 slight slope up to $40^0$ followed by  a sharp cut-off, i.e. no
  photoelectrons
 in the transition region  emit at angles much larger
 than $40^0$ (see fig.2b).
 This remarkable phonomenon corroborates the data 
 of Paulus {\it et al} $[14]$ , which we think is due to the pure tunneling
 nature.
  In contrast, in the mixing regime where multiphoton ionization becomes
  significant, the angular distributions  show no such cut-off,
   and there appears to be emission even at $90^0$.
  Considering the rescattering effects from a simple
  classical model, Paulus {\it et al} find a peak
   at $30^0$ of the angular distribution for
   the electrons in the  transition region.
 However, the emission of the photoelectrons in the direction of the laser
  electric field is largely underestimated $[7]$.
  In our model,
a complete Newtonian
  equation is  used to simulate the electron motion after
  the tunneling so that the effects of multiple returns are included.
Therefore, our model gives more reasonable results on the emission of the photoelectron in field direction.
 Furthermore, the PADs in the regime of plateau exihibit additional peak
  which is  so called side lobe
  phenomenon observed in the experiments.
 One example is shown in  figure 2d for $4U_p\pm 0.1U_p$, there the side lobe occurs at about 
$40^0$.
 The sum of the side lobes for photoelectron in the plateau region are
  responsible for the slight slope followed  by a sharp cut-off
  observed in the Fig.2b.
  Our numerical results also show that the angular distributions both below and
  well above the plateau region are strongly concentrated in the field
  direction.
 Detailed investigations show that,
  aside from these electrons
  which directly drift away without returning to the core ,
    the most electrons in the region I
  experience the forward scattering , while  the electrons
 in the region III are backscattered by almost $180^0$.
 However, in the plateau region the forward scattering and back
 scattering are  equally important and seem have equal probability.

In Fig.3 we demonstrate three typical trajectories of the photoelectrons corresponding to the 
region I, II and III in Fig.1a respectively.
Generally speaking, after tunneling, most  electrons will be driven by the external fields to
return to and interact with the core. This rescattering process greatly determine the final energy and
momentum distributions of photoelectrons. Without considering the rescattering, the 
PADs are expected to become more peaked along the polarization axis as the order of the ATI peak increase,
and the ATI spectrum be a sharply decreasing curve.
However, in the rescattering process,
an electron's energy  changes a lot  through an exchange of momentum
 with the core. It  can obtain high energy
from  a strong collision. Then, a near $180^0$
backscattering which imply a large exchange of momentum certainly gives the highest energy
(region III). A forward scattering with small emission angle  only  provides a relative low energy for 
an electron (region I). Whereas, in the transition region ( region II) the situation become
 quite complicated. Multiple return and long-time trapping can be experienced by those electrons
in this region. The classical trajectories  show  complex behavior. This fact  
leads to the  randomness in emission direction of photoelectrons.
As will be seen later, this is a signature of the chaotic behavior of the trajectories that constitute
the plateau region..
\subsection{Chaotic Behavior in Rescattering Process}
As was shown above, in our model, the rescattering process of the electrons after tunneling can 
be well described by the Newtonian equation (1).
 ATI spectra and PADs can be obtained by calculating statistics on an ensemble of trajectories 
correspinding to different initial field phase and perpendular velocity. 
To investigate the detailed
dynamical mechanism underlying those unusual phenomena such as plateau
structure in ATI data, we shall first fix the velocity and 
calculate
the initial phase ($\phi_0$) dependence  ATI energy and emission angle
(Fig.4,5). As the initial velocity perpendicular
to the polarization of the electric field is relatively large (case a and b),
we have  smooth phase dependence ATI energy and emission angle which imply  regular motion of 
classical trajectories.
In these cases, an ionized electron with lower ATI energy
tends to possess a higher probability (near $\phi_0=0$),
and the plots of angular distribution indicate that hoter photoelectron tends
to  emit in the  polarization direction.
Moreover, in these cases the ATI energy of the photoelectron is smaller than 2$U_p$ approximately.
Considering the exponential form of the expression of ionization rate (13), one can conclude that
these cases
correspond to  a sharply decreasing spectrum such as the region I in Fig.1a. 
Therefore these trajectories have very small contribution to the unusual
distributions in the plateau region in Fig.1a.
For case c and b , things become  quite different.
The dependence of $E_{ATI}$ and the emission angle on the initial phase is poorly resolved in a
region near zero point where many strong peaks are observed.
Successive magnifications (Fig.6) of the
unresolved regions show that any arbitrarily small change in the
 initial phase
may result in a substantial  change of  the final electron energy and emission
angle.
Multiple returns  and even infinite long time trapping can occur in this region. 
This is the character of chaotic behavior, namely, {\em sensitive dependence on the
initial condition},
which has been  observed in many chaotic scattering models $[15]$.
The usual rule
that the higher energy corresponds to  the lower emission angle and the
lower probability
which is identified in the  case a and b
is  broken up now.
In the unresolved regions, some  chaotic trajectories
 can produce rather
high ATI energy ($>2U_p$) and therefore are responsible for  the unusual structure -- plateau in ATI spectra.
Additionally, emission angle of these trajectories are also strange,
this fact leads to the unusual angular distribution in the transition region.
In the unresolved region, the electrons with  chaotic trajectories 
are backscattered or forward scattered randomly,
 this also explains 
the fact observed in the last section , that the coexistence of forward scattering and back scattering
motion of the photoelectrons in the plateau region.

Now we turn to  apply the nonlinear dynamics theory$[16]$ to
 give a quantitative description of the singularity of
 the unresolved regions in the
  phase dependence ATI energy plot. 
We call a value $\phi_0$ a singularity if, in any small
 neighborhood domain
 there is a pair of $\phi_0$  which have different signs in
 their final  moments projected in the field direction, i.e.
 they correspond to the forward scattering or back scattering,
 respectively.
 This singularity results from the unstable manifolds and stable
 manifolds which intersect with each other and form a nonattracting
 hyperbolic invariant set with fractal structure.
Taking the situation of Fig.4c as an example,
 we
  employ the uncertainty exponent technique to obtain the fractal
 dimension of the singular set in the phase axis. We randomly choose many values of $\phi_0$
 in an interval containing the fractal set. We then perturb each value
 by an amount $\epsilon$ and determine whether the final canonical  momentum
 projected in the polarization direction
  corresponding
 to initial $\phi_0, \phi_0-\epsilon$ and $\phi_0+\epsilon$ have the same sign.
 If so , we say that the $\phi$ value  is $\epsilon$-certain; if not,
 we say it is $\epsilon$-uncertain. We do this for several $\epsilon$
 and plot on a log-log scale the fraction of uncertain $\phi_0$ values
 $f(\epsilon)$. The result is plotted in fig.7 which shows a good
 straight line and indicates a power law
 dependence $f(\epsilon) \sim \epsilon^\gamma$,
 where $\gamma = 0.25$.
 The exponent $ \gamma $ is related to the dimension of the fractal
 set of the singular $\phi_s$ values in $\phi_0$-axis in Fig.4c  by $[16]$
 \begin{equation}
 D_0 = 1 - \gamma=0.75    \,\,\, .
 \end{equation}

 As discussed above,  the irregular phase dependence ATI energy and
 emission angle resulting in the unusual phenomena such as
 plateau and side lobe in recorded ATI data,
 come from the chaotic scattering behavior in system (1). To demonstrate 
 phase structure  of the chaotic scattering,
 we choose the initial condition
   as those points in phase space well in the asymptotic region of the
   potential, typically, $x=0, v_x\in(-0.5,0.5)$ and $z=150,
    v_z\in (-2,0)$  and then
   trace the evolution of the 5000 trajectories.
 The stroboscopic map for such an ensemble of scattering trajectories is shown in  
 Fig.8.
  Since 
  the system can flow arbitrarily between the bounded and unbounded regimes,
  it is a non-compact system. The phase sections show a rich and complicated
  dynamical structures  which  confirms the existence of the chaos.
\section{Conclusions and Discussions}
  In this paper, we calculate  the ATI spectra and angular distribution
  in the regime of tunnel-ionization from an improved quasistatic model
  , and show that those unusual phenomena recorded in ATI data can be traced
back to the essential classical behavior of the electrons in the 
combined atom and laser fields. 
  The  properties of ATI spectra and
  angular distribution presented
  in this paper  reflect some fundamental features of
  ionization in the strong field tunneling limit.
  For example, recent high precision measurements of helium photoelectron
  energy and angular distributions in the pure tunnel regime demonstrate
  much similar properties, such as the flat energy distribution that
  extends out to surprisingly high energies before abruptly truncating
   at $8-10U_p$$[13]$.
   In another aspect we associate the unusual phenomena in
   ATI spectra and PADs in a real atom system with the irregular
   chaotic behavior.
    We think that nonlinear dynamics theory is very  helpful in understanding
   the behavior of atoms in the intense laser fields  just like
   its successful applications in photoabsorption spectrum of Rydberg
   atoms$[17]$ and the microwave ionization of the highly excited atoms $[18]$

\section*{Acknowledgments}
This work was supported
 by  the Research Grant Council  and the Hong Kong
Baptist University Faculty Research Grant.
The work done in China was  partially supported
 by the National Natural Science Foundation of China/19674011, the Science Foundation of the 
CAEP and National High Technolofy Committee of Laser . 
We are very grateful to Dr. Baowen Li and Dr. Lei Han Tang
 and all members of the Center for
 Nonlinear Studies for stimulating discussions.

\section*{Captions of Figures}
\begin{itemize}
\item Fig.1 ATI  spectra and total PADs calculated from our model. 
$F = 0.06 a.u., \omega =0.04242 a.u.$.
\item Fig.2 PADs for different energy regions. a,b and c correspond to the region I,II and III in 
fig.1a respectively. The d shows the PAD at energy around $4U_p$, where the side lobe
near  $40^0$ is very obvious.
\item Fig.3 
Three typical trajectaries correspond to the region I, II and III in fig.1a respectively.
a)Initial conditions are $t_0=8.050 a.u., v_{x0}=0.123 a.u.$, the weight $ 2.839\times 10^{-4} a.u$. 
The final ATI energy is $0.29 a.u.$, emission angle $-3.2^{0}$. A weak collision occurs in this case.
b)Initial conditions are $t_0=-6.356 a.u., v_{x0}=0.130 a.u.$, the weight $3.645\times 10^{-4} a.u.$.
The final ATI energy is $3.58 a.u.$, emission angle $-13.3^{0}$. Multiple return occurs in this case.
c)Initial conditions are $t_0=0.871 a.u., v_{x0}=6.00\times 10^{-5} a.u.$, the weight $3.154\times 10
^{-7} a.u.$. The final ATI energy is $5.02 a.u.$, emission angle $-178^{0}$. A strong collision 
occurs in this case.
\item Fig.4 Phase dependence ATI energy spectra  for four different initial
perpendicular velocities.
$F = 0.06 a.u., \omega =0.04242 a.u.$.
\item Fig.5 Phase dependence  PADs for four different initial 
perpendicular velocity.
$F = 0.06 a.u., \omega =0.04242 a.u.$.
\item Fig.6 Successive magnifications of fig4c.
\item Fig.7 Scaling Law of the Singular Set.
\item Fig.8 Stroboscopic map
a) in plane $(x,v_x)$, b) in plane $(z,v_z)$
 for 5000 trajectories originating
asymptotically at $x=0,v_x\in (-0.5,0.5)$ and $z=150, v_z \in (-2,0)$.
\end{itemize}

\newpage
\begin{figure}
\epsfxsize=16cm
\epsffile{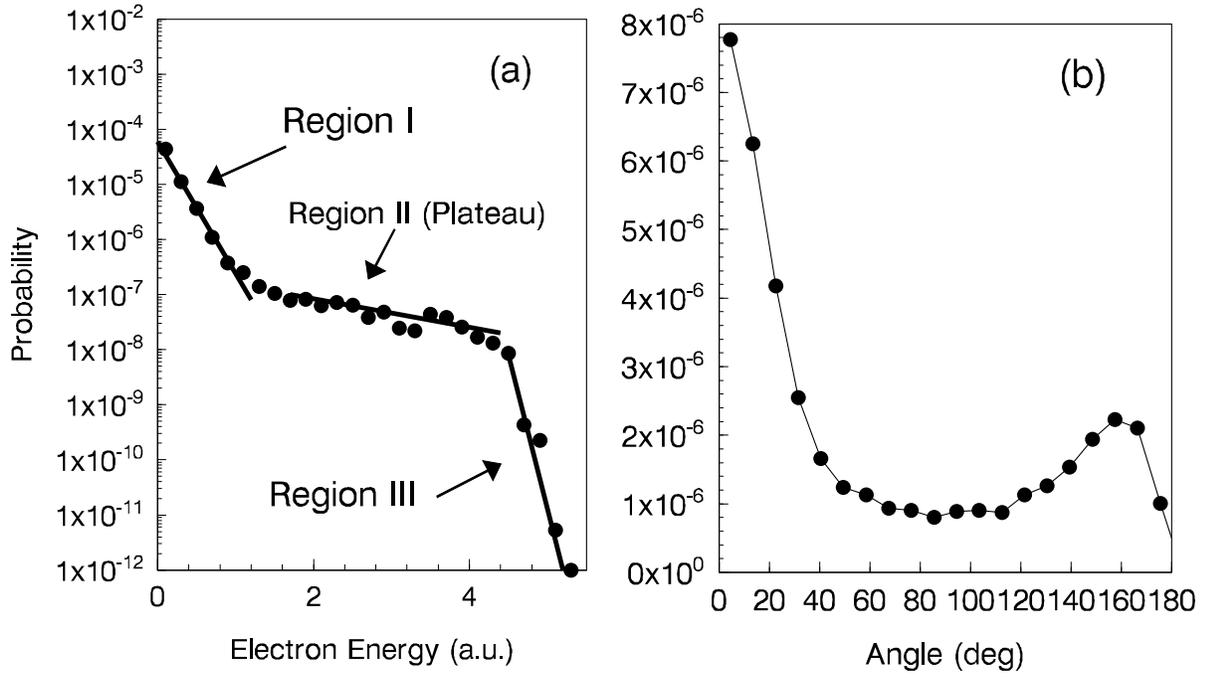}
\smallskip
\caption{
 ATI  spectra and total PADs calculated from our model.
$F = 0.06 a.u., \omega =0.04242 a.u.$.
 }
\end{figure}

\newpage
\begin{figure}
\epsfxsize=12cm
\epsffile{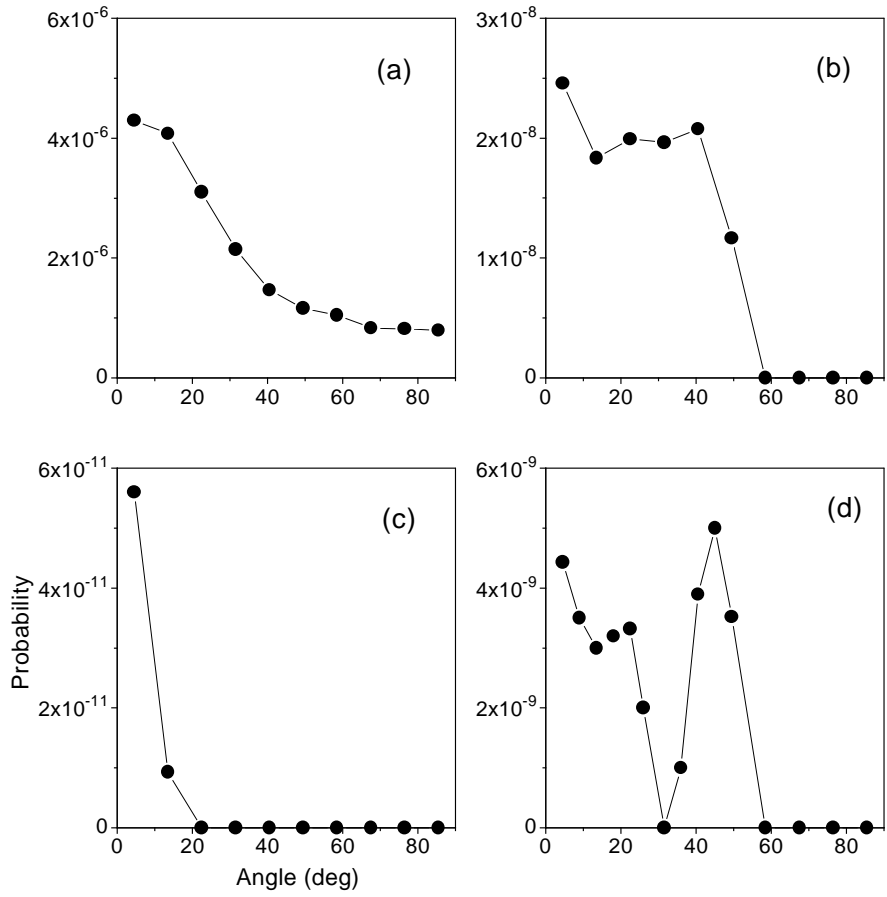}
\smallskip
\caption{
 PADs for different energy regions. a,b and c correspond to the region I,II and III in 
fig.1a respectively. The d shows the PAD at energy around $4U_p$, where the side lobe
near  $40^0$ is very obvious.
}
\end{figure}
\newpage
\begin{figure}
\epsfxsize=12cm 
\epsffile{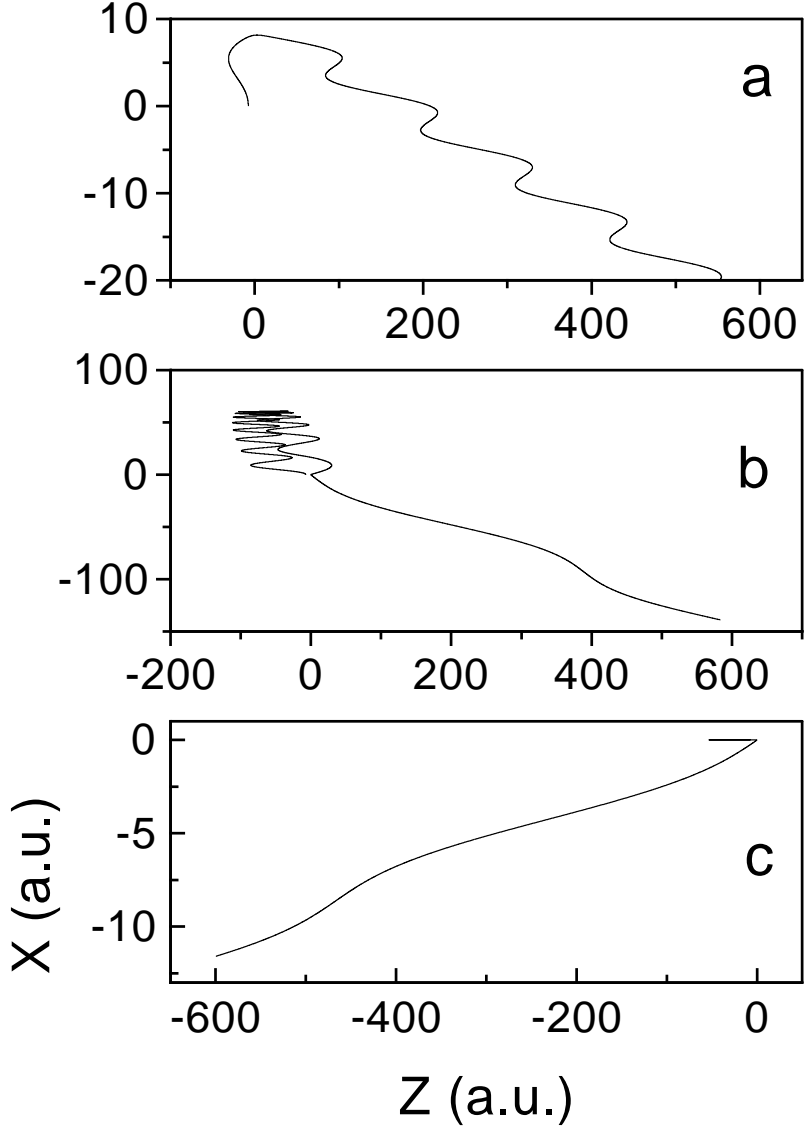}
\smallskip
\caption{
Three typical trajectaries correspond to the region I, II and III in fig.1a respectively.
a)Initial conditions are $t_0=8.050 a.u., v_{x0}=0.123 a.u.$, the weight $ 2.839\times 10^{-4} a.u$. 
The final ATI energy is $0.29 a.u.$, emission angle $-3.2^{0}$. A weak collision occurs in this case.
b)Initial conditions are $t_0=-6.356 a.u., v_{x0}=0.130 a.u.$, the weight $3.645\times 10^{-4} a.u.$.
The final ATI energy is $3.58 a.u.$, emission angle $-13.3^{0}$. Multiple return occurs in this case.
c)Initial conditions are $t_0=0.871 a.u., v_{x0}=6.00\times 10^{-5} a.u.$, the weight $3.154\times 10
^{-7} a.u.$. The final ATI energy is $5.02 a.u.$, emission angle $-178^{0}$. A strong collision 
occurs in this case.
}
\end{figure}

\newpage
\begin{figure}
\epsfxsize=16cm
\epsffile{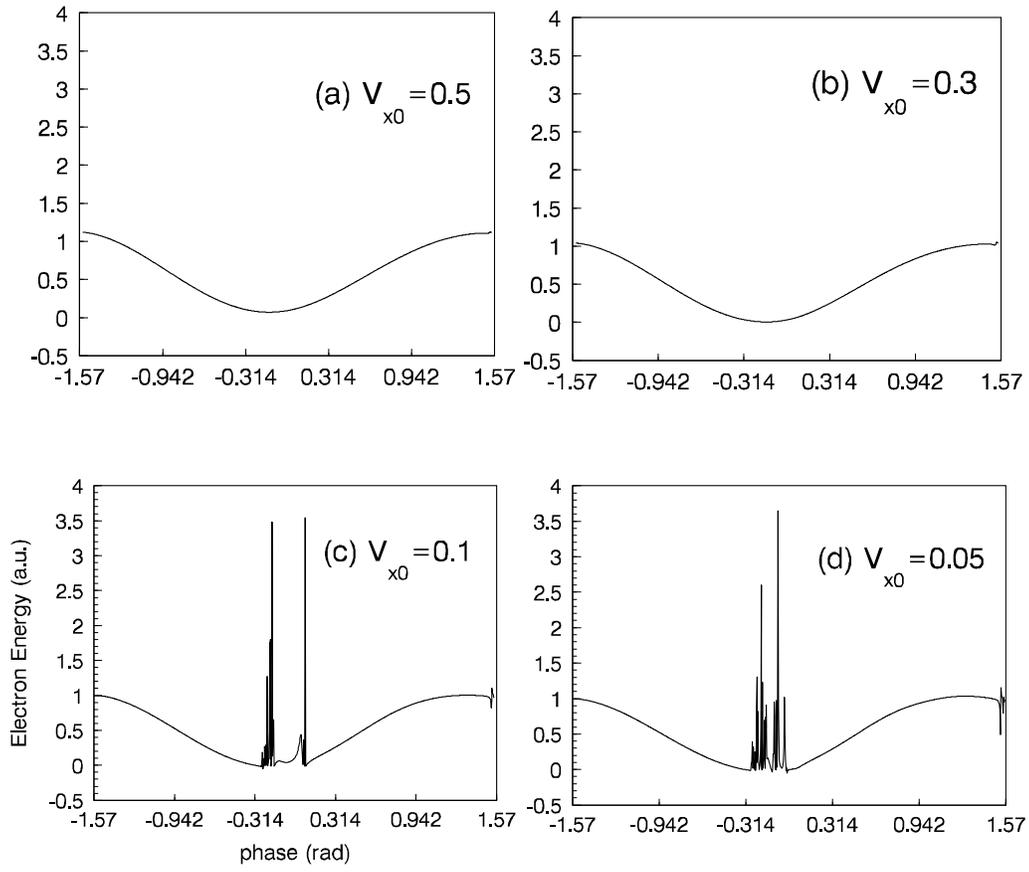}
\smallskip
\caption{
 Phase dependence ATI energy spectra  for four different initial
perpendicular velocities.
$F = 0.06 a.u., \omega =0.04242 a.u.$.
}
\end{figure}
\newpage
\begin{figure}
\epsfxsize=16cm
\epsffile{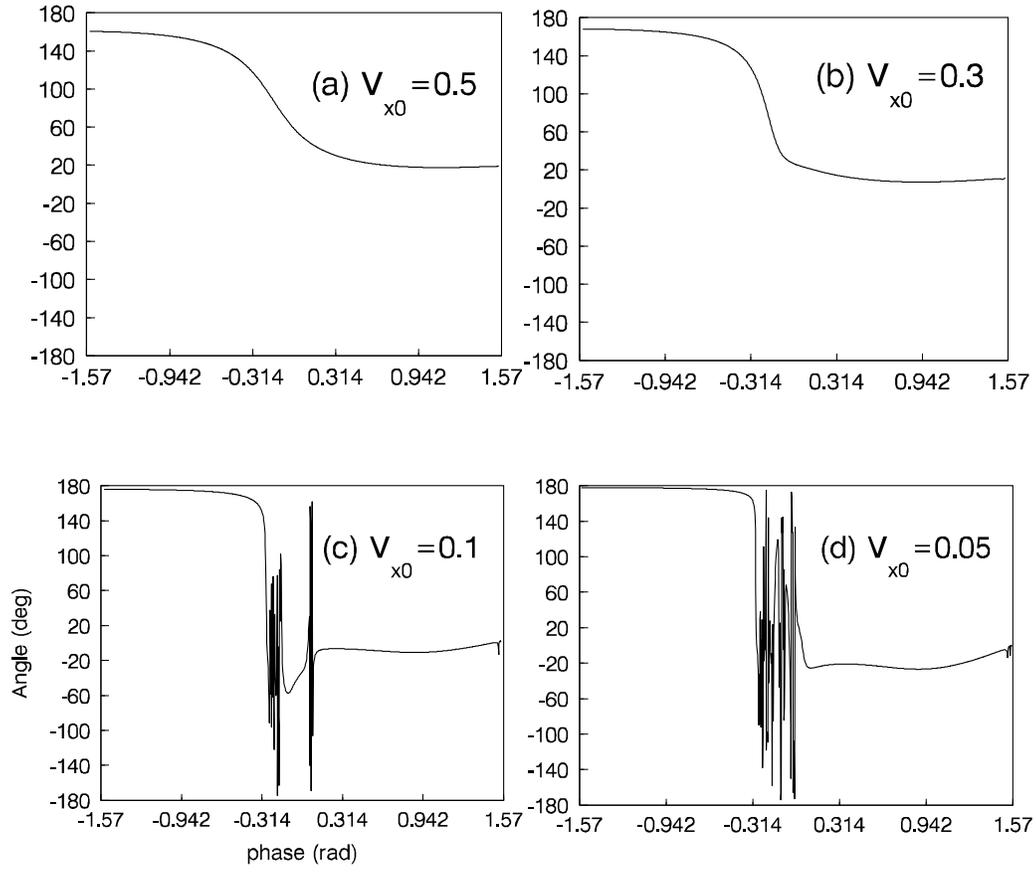}
\smallskip
\caption{
 Phase dependence  PADs for four different initial 
perpendicular velocities.
$F = 0.06 a.u., \omega =0.04242 a.u.$.
}
\end{figure}

\newpage
\begin{figure}
\epsfxsize=16cm
\epsffile{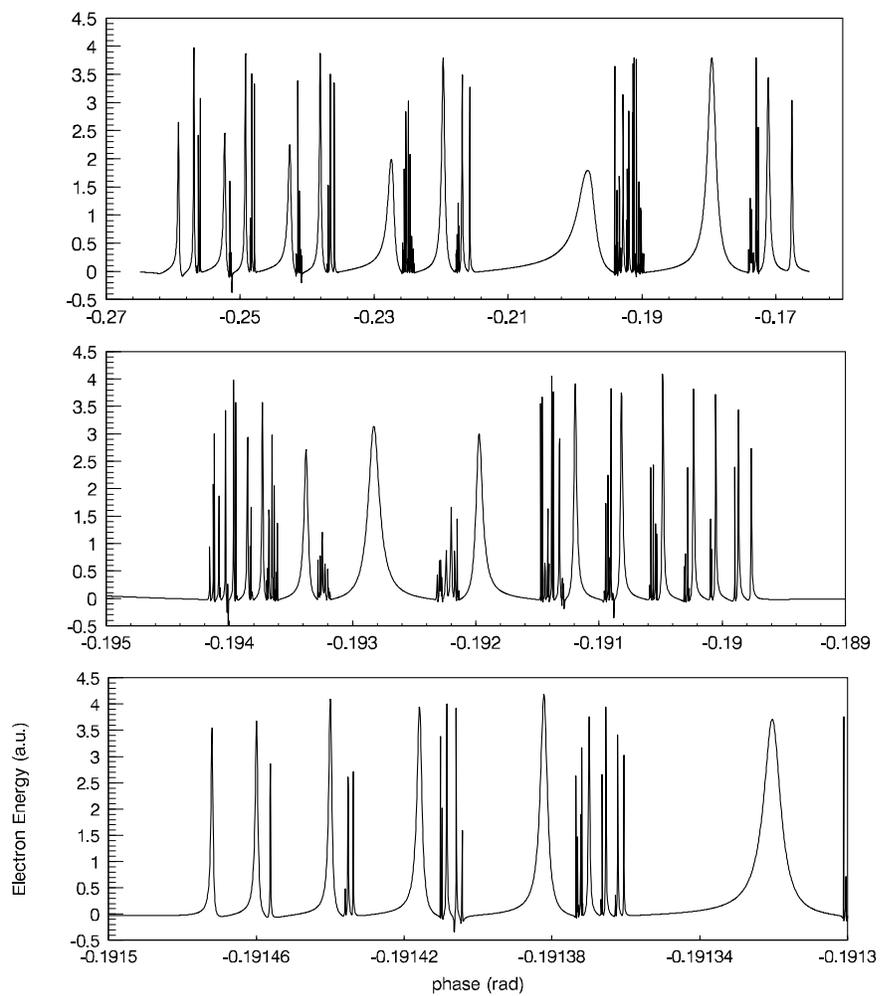}
\smallskip
\caption{
 Successive magnifications of fig4c.
}
\end{figure}
\newpage
\begin{figure}
\epsfxsize=16cm
\epsffile{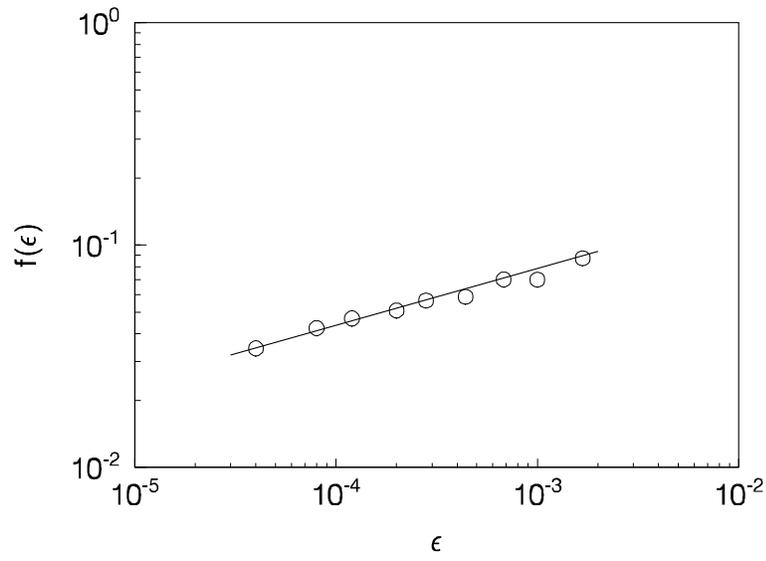}
\smallskip
\caption{    
 Scaling Law of the Singular Set.
}

\end{figure}

\newpage
\begin{figure}
\epsfxsize=12cm
\epsffile{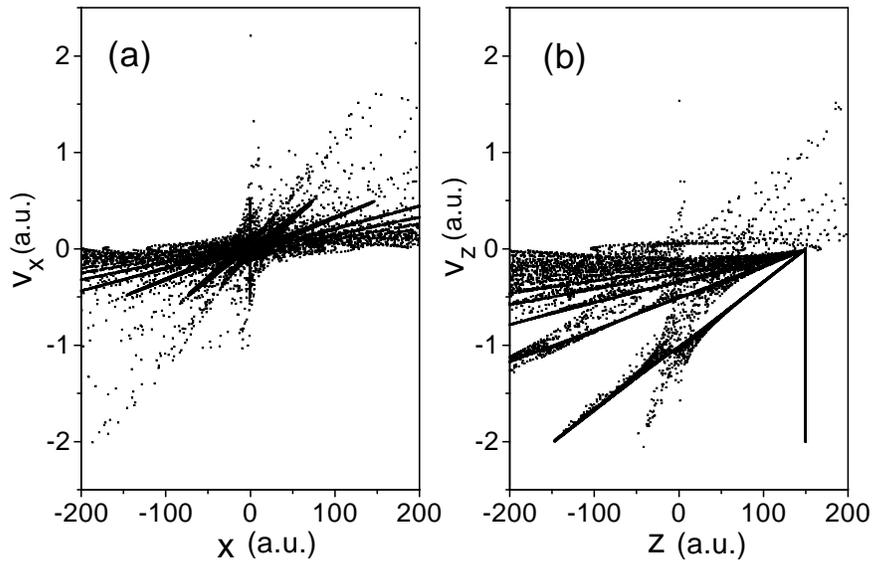}
\smallskip
\caption{
Stroboscopic map
a) in plane $(x,v_x)$, b) in plane $(z,v_z)$
 for 5000 trajectories originating
asymptotically at $X=0,v_x\in (-0.5,0.5)$ and $Z=150, v_z \in (-2,0)$.
}
\end{figure}
\end{document}